# Polymorphic gene conferring susceptibility to insulin-dependent diabetes mellitus typed by ps-resolved FRET on nonamplified genomic DNA

Luca Nardo [1,2], Giovanna Tosi [3], Maria Bondani [4], Roberto S. Accolla [3], Alessandra Andreoni [1,2]

[1]Department of Science and High Tech, and [2]C.N.I.S.M. (Consorzio Nazionale Interuniversitario per le Scienze Fisiche della Materia), University of Insubria, Via Valleggio 11, 22100 Como, Italy, [3] Department of Surgical and Morphologic Sciences, University of Insubria, Viale Borri 57, 21100 Varese, Italy, [4] Institute for Photonics and Nanotechnology, C.N.R. (Consiglio Nazionale delle Ricerche), Via Valleggio 11, 22100 Como, Italy.

Corresponding author:   Alessandra Andreoni
                        Department of Science and High Tech, and C.N.I.S.M., University of Insubria
                        Via Valleggio 11, 22100 Como, Italy
                        Alessandra.Andreoni@uninsubria.it

*Classification:*
PHYSICAL SCIENCES: Applied Physical Sciences
BIOLOGICAL SCIENCES: Immunology

ABSTRACT

This work concerns the identification of the allelic sequences of the DQB1 gene of the human leukocyte antigen system conferring susceptibility to the development of insulin-dependent diabetes mellitus (IDDM) in DNA samples with no need of PCR amplification.

Our method is based on the time-resolved analysis of a Förster energy-transfer mechanism that occurs in a dual-labeled fluorescent probe specific for the base sequence of the allelic variant of interest. Such an oligonucleotide probe is labeled, at the two ends, by a pair of chromophores that operate as donor and acceptor in a Förster resonant energy-transfer. The donor fluorescence is quenched with an efficiency that is strongly dependent on the donor-to-acceptor distance, hence on the configuration of the probe after hybridization with the DNA containing or not the selected allelic sequence. By time-correlated single-photon counting, performed with an excitation/detection system endowed with 30-ps resolution, we measure the time-resolved fluorescence decay of the donor and discriminate, by means of the decay-time value, the DNA bearing the allele conferring susceptibility to IDDM from the DNAs bearing any other sequence in the same region of the DQB1 gene.

## INTRODUCTION

The human leukocyte antigen (HLA) system encodes highly polymorphic molecules which are fundamental for the triggering of the immune response. Moreover, it is known that the expression of certain HLA molecules is associated with the susceptibility to many autoimmune diseases, including insulin-dependent diabetes mellitus (IDDM) (1, 2). Distinguishing the subjects with susceptibility genes, before the onset of the disease, calls for the development of rapid high-throughput HLA typing technologies. On one hand, such an ambitious aim seems to contrast with the increasing number of susceptibility conferring alleles that are being discovered in patients affected by specific pathologies (3). On the other hand, it is encouraging that a massively parallel sequencing technology has recently allowed mutation detection in picoliters of cancer specimens (4). For large-scale HLA analysis, we decided to set a method allowing the typing of genomic DNA simply purified from leukocytes without the need of polymerase chain reaction (PCR) amplification. We adopted a DNA probe matching the sequence of one of the most common alleles conferring susceptibility to IDDM in Caucasian population and, by making a novel use of Förster resonant energy transfer (FRET), we found a method for detecting sequence variations of even a single nucleotide in DNA template used as target. Markers of susceptibility to IDDM have been recognized in the HLA-II region and, in particular, in HLA-II DQA1 and DQB1 loci, coding for the α and β chains of DQ heterodimer, respectively (5). Extensive molecular studies of HLA-DQ genotype in patients with IDDM led to correlate the susceptibility with the polymorphism of a single aminoacid in both DQα and DQβ chains (6, 7). In particular, the combined presence of an arginine residue in position 52 of DQα chain and of a neutral residue (serine, alanine or valine) in position 57 of DQβ chain is strongly associated with diabetes onset. On the contrary, the presence of aspartic acid at position 57 of DQβ confers resistance to IDDM (2).

In the present work we focussed on the typing of DQB1 locus by hybridizing target DNAs to a single probe corresponding to DQB1 0201 "susceptible" allele, which was dual labelled by a fluorescent reporter in 5' and a non-fluorescing quencher in 3'. In such a probe, the fluorescence of the donor tetramethylrhodamine (TAMRA) is quenched via FRET by the black-hole quencher (BH2) depending on the conformational differences among the hybrids due to the probe/target mismatches. It is worth noting that fluorescence quenching by FRET is basic to many genotyping methods that include a step in which allele-specific hybridization products must be recognized. These methods rely on the fact that FRET removes the excitation energy deposited on the donor (D) towards the acceptor (A) with an efficiency that depends on the D-to-A distance. As a result, when D and A are in close proximity, the D fluorescence is weaker than that emitted when they are at greater distances from each other. Multiple allele-specific probes, each labelled with a different D-A pair, can be used for querying allelic sequences (8). The sequence of each probe is complementary to a specific polymorphic sequence and probes are mixed all at once with the PCR-amplified template. The highly stringent hybridization conditions allow annealing only of the perfectly matching probe, which results in a greater D-to-A distance, thus in an enhanced fluorescence emission of D. Conversely, as the non-matching probes preserve the native coiled conformation, their fluorescence is quenched. The genotype of the template DNA is inferred by assessing which D is the fluorescing one. We exploit the fact that FRET can produce quantitatively different fluorescence emissions depending on the probe-target hybrid conformation, thus making it feasible SNP genotyping by means of a single probe. Pioneering attempts in this direction have been made by Gaylord and collaborators (9) and, later, by Al Attar and collaborators (10). The latter group used a cationic polymer conjugated to the targets as D and monitored the fluorescence intensity of A attached to a single peptide nucleic acid probe. As the neutral peptide backbone of the probe did not bind to D, the acceptor A received the excitation energy only upon hybridization of the probe to the targets. By applying this scheme to typing of ABL polymorphism in the BCR-ABL oncogene, Al Attar et al. have determined significantly different values of the FRET efficiency between D and A when the probe was hybridized to each of the possible target sequences (*i.e.* to each of the possible mutated alleles). Thus, they demonstrated the feasibility of ABL typing by using a single probe. It is easy to realize that distinguishing among the hybrids that our probe forms with the different allelic sequences by absolute measurements of D-fluorescence intensity is a hopeless approach to the development of a typing method suitable for non-amplified genomic DNA. Indeed, a high-throughput screening method based on FRET must work with cellular lysates, where the DNA concentration is variable from sample to sample, and utilize detectors whose response is stable against environmental stray light conditions. In our approach we quantify the FRET efficiencies of each probe/template hybrid by analyzing the D fluorescence temporal decay on a very fast time scale. We measure the TAMRA-fluorescence decay time ($\tau_D$) upon picosecond laser excitation, at a wavelength that is only absorbed by this chromophore, and detect the fluorescence by a time-correlated single-photon counting apparatus, which is endowed with a temporal resolution of <30 ps. Though the acquisition rate is less than one photon per excitation pulse, which ensures the applicability of our technique to the detection of even few probe-target duplexes, good determinations of the $\tau_D$ values are done in few minutes due to the megahertz repetition rate of the excitation laser. The slight differences in the distance of the BHQ2 acceptor from the TAMRA donor and the consequent differences in FRET efficiencies produce distinct $\tau_D$ values for the probe hybridized to each of the allelic sequences.

## RESULTS

### Identification of allele-specific HLA-II DQB1 oligonucleotides

In order to establish our FRET-based HLA-II DQB1 typing method we first set the system by using a panel of synthetic oligonucleotides listed in **Fig. 1.** These oligonucleotides are complementary to the coding sequences of the eight relevant DQB1 alleles and encompass codons 51 to 60 (6). Nucleotide variations with respect to the 0201 allele, shown on the top, are marked. Each oligonucleotide was hybridized with the probe corresponding to the 0201 variant having the GCC codon for alanine in position 57. The oligoprobe and the structures of the TAMRA donor and BHQ2 acceptor molecules are shown in **Fig. 2**. In general, to increase the specificity of a probe for its target DNA, it is desirable to design the oligonucleotide with a great rather than a small number of nucleotides. However, this is not true for probes dual-labeled at the 5'- and 3'-positions by a D and an A interacting by FRET, in that the D-to-A distance cannot remarkably overcome the Förster radius. We designed a probe of 22 nucleotides covering from codon 51 up to the first base of codon 58 as we preliminarily demonstrated that probe/target mismatches next to the probe ends, where D or A are covalently bound, cause the greatest changes in the FRET efficiency (11, 12). In the present case, mismatches at the polymorphic codon 57, which occur in the cases of all variants except 0201 and 0302, are expected to prevent pairing of the last three-four bases on the side of the BHQ2. Such an "unzipping" of the duplex structure should let A approach D and cause a sizeable decrease in the donor-fluorescence decay time ($\tau_D$). It is worth noting that the eight DQB1 alleles are all identical from codon 58 to codon 60, thus further extending the probe would prevent unzipping.

The $\tau_D$ values that we determined for the DQB1 0201 probe binding to each of the eight synthetic oligonucleotides are listed in column a) of **Table 1**. For all hybrids, except those that the probe forms with the variant sequences 05031 and 0602, we find significantly different $\tau_D$ values. Most importantly, the greatest value corresponds to the perfect matching of the probe with the oligonucleotide having the sequence of the DQB1-0201 allele. The next highest $\tau_D$ value is exhibited by the probe binding the 0302 sequence, which is the only one displaying a base composition identical to that of the 0201 allele at codon 57.

The manner in which the other values of $\tau_D$ scale agrees with what we previously observed by using a probe containing 18 instead of 22 nucleotides (12). Moreover, as the 18-nucleotide probe did not cover codon 51, which is identical in all allelic variants, the unzipping of the 5'-end observed for the 18-nucleotides probe does not occur here. Thus, as shown in **Table 1**, for the mismatching oligonucleotides mimicking the sequences of the DQB1 allelic variants of other than 0201, we measure $\tau_D$ values that are both greater than those for the 18-nucleotide probe (see the cases of alleles 0501, 0502, 05031, and 0602, which produced very fast decays when hybridized to the 18-nucleotide probe) and much more similar to that measured for the perfectly matching target oligonucleotide.

For these target sequences we had hypothesized an unzipping including the first 5 nucleotides, due to the double mismatch of the second and fifth nucleotides. However, addition of the three nucleotides corresponding to codon 51 confers a greatly enhanced selectivity to the 22-nucleotide probe and does not prevent discrimination of the DQB1-0201 genotype. Beside the above considerations, the scaling of the values in column a) of **Table 1** agrees with the explanations that we presented in the quoted work of the dependence of the $\tau_D$ changes on both number and position of the base mismatches (12). Moreover, the data obtained in the present work with the 22-nucleotides probe, which we deemed ideal for typing of genomic DNA templates, confirm that the type of mismatch can also influence the decay time. In fact, the 0501 and 0602 alleles, having only one nucleotidic variation in the same position of codon 57 (CAA and CTA, respectively) display significantly different $\tau_D$ values when hybridized to the 0201 probe.

### HLA-II DQB1 typing of non-amplified genomic DNA

In order to validate our FRET-based HLA-II DQB1 typing approach set with synthetic oligonucleotides, we tested the system with non PCR-amplified genomic DNAs. DNA was extracted from five HLA-homozygous lymphoblastoid cell lines having the following DQB1 specificities: PITOUT (0201); HOM-2 (0501); JVM (0301); WT51 (0302); OLL (0402) (6). Purified double-stranded DNA was denatured by heating and then hybridized to the dual-labeled DQB1 0201 probe at a specific annealing temperature as indicated in Methods. Upon suitable modifications of the setup as to the excitation/collection optics, the corresponding $\tau_D$ values were measured (column b of **Table 1**). These values are surprisingly equal to those obtained with the oligonucleotides corresponding to the same DQB1 allele, which are listed on the same lines in column a), in spite of the fact that these measurements were performed on entire genomic DNA.

## DISCUSSION

In this study we successfully developed a new method to identify the DQB1-0201 variant, an allele that is statistically over-represented in patients with insulin-dependent diabetes mellitus (type-1 diabetes). Indeed this allele is considered a susceptibility marker for people predisposed to type-1 diabetes. The method is based on measurements of fluorescence decay times performed with single photon sensitivity and picosecond resolution. Purified, non-amplified genomic DNAs bearing different sequences in the polymorphic region of the HLA-DQB1 gene were hybridized with a single

DNA probe corresponding to the DQB1-0201 allele. The probe was dual-labelled with fluorescence donor-and-acceptor chromophores at its opposite ends. The chromophores undergo FRET, whose efficiency is characterized by measuring the fluorescence decay time $\tau_D$ of the donor. As the FRET efficiency, and hence the $\tau_D$ value, is differently modified by the number of probe-to-target mismatches and by their positions, the hybrids formed by the probe with each allelic variant are identified by distinct values of $\tau_D$. It is worth noting that the picosecond-laser excitation wavelength we adopted, 532 nm, is very close to the absorption peak of the TAMRA donor, which falls at 556 nm, being not even absorbed by the non-fluorescent BHQ2 acceptor.

To determine the values of $\tau_D$ we devised suitable strategies for taking into account that, though the excitation pulses are only absorbed by TAMRA, not all the TAMRA molecules that contribute to the detected fluorescence-decay patterns undergo FRET and decay with time-constant $\tau_D$. In both oligonucleotide and DNA samples, there can be dual-labelled fluorescent probes left free in solution, thus keeping their natural conformation as single-stranded probe and contributing decay-components shorter than $\tau_D$. There might also be probes lacking the BHQ2 acceptor as well as TAMRA molecules not bound to the probe: both of them decay slower than any TAMRA quenched via FRET by the presence of a more-or-less distant BHQ2. The overall effect is that the rough experimental decay patterns deviate from single exponentials. In the case of oligonucleotides the decays contain fast transients of negligible amplitude, which become irrelevant after <800 ps from the peak of the fluorescence signals. This fact not only allows a simple analysis of the experimental decays, but also reveals the almost complete annealing of the dual-labelled probe molecules to the target oligonucleotides. In the case of DNAs the probe binding to the target sequences is contrasted by the presence in solution of the complementary sequences and by partial annealing to genomic DNA regions other than the allelic-variant under examination. However, a neutral application to all experimental decay patterns of the strategy described in Methods and already used in a previous work of ours (12), brought to decays ($D_{xxx}$) of remarkable amplitudes that could be fitted by the sum of two exponentials. The faster one has a decay time $\tau' = 709 \pm 67$ ps and is attributed to the partially annealed probes, while the slower one decays with the $\tau_D$ values reported in **Table 1**.

The HLA typing method reported is relatively simple; the instrumentation that is needed is not particularly expensive and is prone to be adapted for routine use in a diagnostic laboratory. The great advantage of processing very quickly a considerable number of samples and the high specificity of the detection may allow the use of this method not only for routine typing of HLA, but also for the typing of specific "pathologic" sequences, for example of mutated oncogenes, in order to identify the specific mutation in tumor tissues and most importantly the extent of mutated genes in pre-neoplastic lesions.

## METHODS

**Hybridization of oligonucleotides.** The D-A pair in the single-stranded dual-labelled oligonucleotide probe, D-5'(ACGCTGCTGGGGCTGCCTGCCG)3'-A, was the tetramethylrhodamine analogue 5'-Dimethoxytrityloxy-5-[N-((tetramethylrhodaminyl)-aminohexyl)-3-acrylimido]-2'-deoxyUridine-3'-[(2-cyanoethyl)-(N,N-diisopropyl)]-phosphoramidite (TAMRA) and the black hole quencher 4'-(4-Nitro-phenyldiazo)-2'-methoxy-5'-methoxy-azobenzene-4''-(N-ethyl)-N-ethyl-2-cyanoethyl-(N,N-diisopropyl)-phosphoramidite (BHQ2). The dual-labeled probe, purified by HPLC and the oligonucleotides were purchased from TIB Molbiol (Genova, Italy). The lyophilized samples, each one in an amount of about 20 nM, were diluted in Tris-HCl/EDTA buffer (pH 7.6; 10 mM ionic strength) down to concentration values that were determined spectrophotometrically from UV absorbance spectra over 1-cm pathway. The probe and each target oligonucleotide were diluted to the same final concentration and mixed in equal volumes (final concentration: 250 nM). The mixtures were gradually heated at a rate of +3 °C/min to 98 °C and kept at this temperature for 10 min to allow for the denaturation of secondary structures of the single strands and to destroy the dimers that could be present. The solutions were then cooled down, at a rate of −1 °C/min, to a plateau temperature of 73 °C, which is lower by 1 °C than the melting temperature of the hybrids. The samples were kept at the plateau temperature for 20 min, in order to maximize annealing. It is worth noting that the hybrid melting temperatures are different for hybrids formed by the dual-labeled probe with the different oligonucleotides: the value is 74 °C for the DQB1-0201 oligonucleotide (TIB Molbiol, private communication) but, for the other hybrids, it is lower due to the base mismatches. In an attempt to favor the probe annealing also to the other oligonucleotides, we adopted a plateau temperature of 67 °C in the case of 0302 (two base mismatches), 64 °C in the case of 0301 (three base mismatches), 61 °C in the cases of 05031 and 0402 (four base mismatches), 58 °C for 0501, 0502 and 0602. In practice, starting from the 73 °C value used as the plateau temperature for the probe annealing to the perfectly matching 0201 variant, we adopted values three degrees lower for every additional mismatch. Finally the hybridized solutions were slowly brought to room temperature (−2 °C/min) to be measured with the TCSPC apparatus. The whole procedure of temperature-controlled annealing was carried out in the thermostat of a ThermoQuest-Finnigan gas chromatograph (San Jose, CA, US).

**Hybridization of genomic DNA.** Five lymphoblastoid cell lines, HLA homozygous (13), were used to extract genomic DNA by standard procedures. The procedure of temperature-controlled annealing performed after DNA denaturation at 98 °C was similar to that adopted for oligonucleotides, except for the fact that the five samples were put together in the thermostat and exposed to a sequence of five 20-min periods in which the plateau temperatures were set, in order, at the values 73 °C, 67 °C, 64 °C, 61 °C, and 58 °C. These decreasing temperature values should allow optimal annealing of the probe to the DQB1 antisense alleles in the following order: 0201 first, then 0302, 0301, 0402, and 0501 at last.

**Measurements of D-fluorescence decay time.** All TCSPC measurements were performed at room temperature. In the case of oligonucleotides the fluorescence was detected at 90 deg with respect to laser excitation and the sample was contained in 1 cm × 1 cm quartz cuvette with four polished windows. In the case of DNAs, due the smallness of the available sample volumes, we adopted an epi-fluorescence 20× microscope geometry in which the wells containing 200 µl of the samples were excited and the fluorescence collected along the same direction. The D-fluorescence was separated from the reflected/back-scattered excitation by a dichroic mirror at 45 deg (DMLP567, Thorlabs GmbH, Dachau, Germany). The excitation/detection apparatus, extensively described in a previous work of ours (12), is entirely made up of commercial components. Its main features are: excitation pulses at 532 nm (6.4 ps duration at 113 MHz repetition rate), typically attenuated by a factor >$10^3$ to achieve single-photon regime in the fluorescence detection; single-photon avalanche diode with built-in active quenching circuitry and thermoelectric cooling as the detector; fluorescence decay patterns recorded with 2.44 ps/channel resolution; full-width at half-maximum duration of the detected excitation pulse of <30 ps; time-to-amplitude conversion operated in reversed start-stop modality so that each detected photon produced a valid start signal; fluorescence detected above 550 nm; each decay pattern acquired up to a fixed number of counts at the peak, namely 65535; each acquisition time recorded; each measurement repeated three times at least. The zero of the time scale was set at the peak channel for all decay patterns.

In addition to the measurements on the hybrids, we also measured the fluorescence decay of a solution of the pure single-stranded probe at the same concentration as in the hybridized samples. Preliminarily to data analysis, the rough decay patterns acquired for both this solution and those containing the hybridized oligonucleotides were normalized to the acquisition time and to the absorbance values at 556 nm (TAMRA peak). The rough decay data were treated as described below.

**Data analysis for $\tau_D$ determination: oligonucleotides.** All decay patterns exhibit fast initial transients of negligible amplitudes, thus showing that dual-labelled fluorescent probes left free in solution are virtually absent in our oligonucleotide samples. Such a fast transient represents the majority of the fluorescence decay of the solution of the pure single-stranded probe, in which it is summed to the slow decay component emitted by free TAMRA. Thus, after normalizing to acquisition time and TAMRA absorbance, we subtract this slow component (with the amplitude and lifetime detected for the pure single-stranded probe, see below) from all decays and fit the difference starting from 400 to 800 ps after the peak. For all hybridized oligonucleotide samples we find very good fittings by using single exponentials whose time constants we interpret as the $\tau_D$ values (see first column in **Table 1**). As an example we show the case of the oligonucleotide mimicking the DQB1-0201 in **Fig. 3**.

**Data analysis for $\tau_D$ determination: DNAs.** We apply a modified version of the procedure described in a previous work of ours (12). As in all measured decays patterns, preliminarily normalized to acquisition time and TAMRA absorbance, we recognize a slow decay component typical of not-quenched TAMRA, either free in solution or bound to hybridized probes lacking the BHQ2 acceptor, we begin by subtracting this component from the rough experimental decays of all the hybridized DNA samples with the initial amplitude and lifetime that we detected in the decay pattern measured on the solution of pure probe. The resulting patterns are normalized to their integrals. Let us call these normalized decay patterns $F_{201}$, $F_{501}$, …, shortly $F_{xxx}$. The subtraction is performed also on the probe decay pattern, which is then normalized to its integral to obtain $F_{PROBE}$. We then observe that $F_{PROBE}$ is well fitted by a three-exponential decay

$$Y_{PROBE}(t) = A_{1P} \exp(-t/\tau_{1P}) + A_{2P} \exp(-t/\tau_{2P}) + A_{3P} \exp(-t/\tau_{3P}) \qquad (1)$$

The best procedure we found to get rid of the $F_{PROBE}$ contributions in the decays $F_{201}$, $F_{501}$, $F_{301}$, $F_{302}$, and $F_{402}$, of the samples containing the DNAs was as follows: we carry out multi-exponential fits of the latter decays from which we only extract the values $A_{1xxx}$, $A_{2xxx}$, and $A_{3xxx}$ of the three fastest components and use them to perform the following weighted subtractions

$$F_{xxx} - (A_{1xxx} + A_{2xxx} + A_{3xxx})/(A_{1P} + A_{2P} + A_{3P}) F_{PROBE} \equiv D_{xxx} \qquad (2)$$

that yield the final curves $D_{201}$, $D_{501}$, $D_{301}$, $D_{302}$, and $D_{402}$. Curves $D_{xxx}$ should only contain the fluorescence arising from the desired double-stranded fraction of each sample with decay time $\tau_D$ and from the probe molecules partially annealed to other regions of the genomic DNA target. Actually, all curves $D_{xxx}$ had dynamic ranges covering more than one decade and exhibited bi-exponential decays starting from ~500 ps after the peak. Best fits of the $D_{xxx}$ curves provide the values of the corresponding decay times $\tau'$ and $\tau_D$. The whole procedure is illustrated in **Fig. 4** for the DNA of the cell line homozygous for the DQB1-0201 allelic variant.

**ACKNOLEDGEMENTS**

We thank A. Maspero for making the Finnigan instrument available and N. Camera for his technical assistance. We also acknowledge the funding of the equipment maintenance that is granted by Centro Servizi e Grandi Attrezzature per lo Studio e la Caratterizzazione della Materia of the University of Insubria. LN position is funded by project POR FSE Regione Lombardia, Ob.2 Asse IV 2007-2013, through an FP7 cooperation programme. This work was supported by the following grants to RSA: Fondazione Cariplo 2008-2230 "Cellular and molecular basis of human retroviral-dependent pathology"; A.I.R.C IG 8862 ""New strategies of tumor vaccination and immunotherapy based on optimized triggering of anti-tumor CD4+ T cells";  MIUR-PRIN project 2008-WXF7KK " New strategies of immunointervention against tumors". AA and GT also acknowledge support by University of Insubria through projects  "FAR 2009" and "FAR 2010" .

**REFERENCES**

1. Nepom GT (1990) HLA and type I diabetes. *Immunol Today* 11: 314-315.

2. Tosi G, Facchin A, Pinelli L, Accolla RS (1994) Assessment of the DQB1-DQA1 complete genotype allows best prediction for IDDM. *Diabetes Care* 17: 1045-1049.

3. Tipu HN, Ahmed TA, Bashir MM (2011) Human leukocyte antigen class II susceptibility conferring alleles among non-insulin dependent diabetes mellitus patients. *J Coll Phys Surg Pak* 21: 26-29.

4. Thomas RK, et al. (2006) Sensitive mutation detection in heterogeneous cancer specimens by massively parallel picoliter reactor sequencing. *Nat Med* 12: 852-855.

5. Todd JA, Bell JI, McDevitt HO (1987) HLA-DQB gene contributes to susceptibility and resistance to insulin dependent diabetes mellitus. *Nature* 329: 599-604.

6. Tosi G, et al. (1993) HLA-DQB1 typing of north-east Italian IDDM patients using amplified DNA, oligonucleotide probes, and a rapid DNA-enzyme immunoassay (DEIA). *Molec Immunol* 30: 69-76.

7. Tosi G, et al. (1994) The complex interplay of the DQB1 and DQB2 loci in the generation of the susceptibility and protective phenotype for insulin-dependent diabetes mellitus. *Molec Immunol* 31: 429-437.

8. Tyagi S, Bratu DP, Russell Kramer F (1998) Multicolor molecular beacons for allele discrimination. *Nat Biotechnol* 16: 49-53.

9. Gaylord BS, Massie MR, Feinstein SC, Bazan GC (2005) SNP detection using peptide nucleic acid probes and conjugated polymers: applications in neurodegenerative disease identification. *Proc Natl Acad Sci USA* 102: 34-39.

10. Al Attar HA, Norden J, O'Brien S, Monkman AP (2008) Improved single nucleotide polymorphysm detection using conjugated polymer/surfactant system and peptide nucleic acid. *Biosensors & Bioelectronics* 23: 1466-1472.

11. Andreoni A, Bondani M, Nardo L (2009) Feasibility of single nucleotide polymorphism genotyping with a single-probe by time-resolved Förster resonance energy transfer. *Mol Cell Probes* 23: 119-121.

12. Andreoni A, Bondani M, Nardo L (2009) Time-resolved FRET method for typing polymorphic alleles of the human leukocyte antigen system by using a single DNA probe. *Photochem Photobiol Sci* 8: 1202-1206.

13. Yang SY, Milford E, Dupont B (1989) in *Immunology of HLA 1*, ed. Dupont B. (Springer, New York). pp. 11-19.

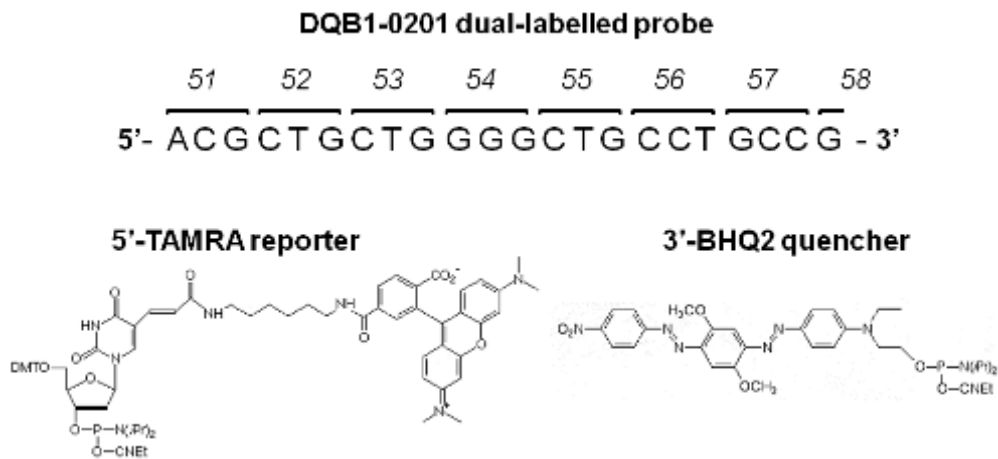

**Figure 1**
Sequences (antisense) of the region comprising codons 51 to 60 of the eight relevant DQB1 allelic variants (listed on the left). Nucleotide changes with respect to DQB1-0201, are marked. The regions shaded in grey are those not covered by the oligonucleotide used as a probe (see **Fig. 2)**.

**Figure 2**
FRET probe matching the DQB1-0201 sequence between codons 51 and 58 (22 bases) and carrying the TAMRA donor and the BHQ2 quencher, whose structures are shown in the bottom.

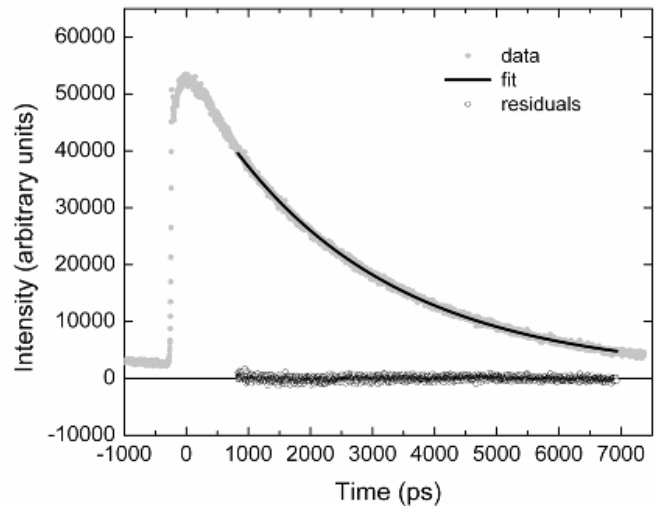

**Figure 3**
Fluorescence decay of the TAMRA donor of the probe bound to the oligonucleotide mimicking the sequence of the DQB1-0201 allele. Grey dots: experimental data after subtraction of the slow decay of TAMRA molecules left free in solution. Black line: fitting curve with best-fitting parameters $\tau_D$ = 2722.61 ps and 652.3 counts as the background. Open dots: residuals.

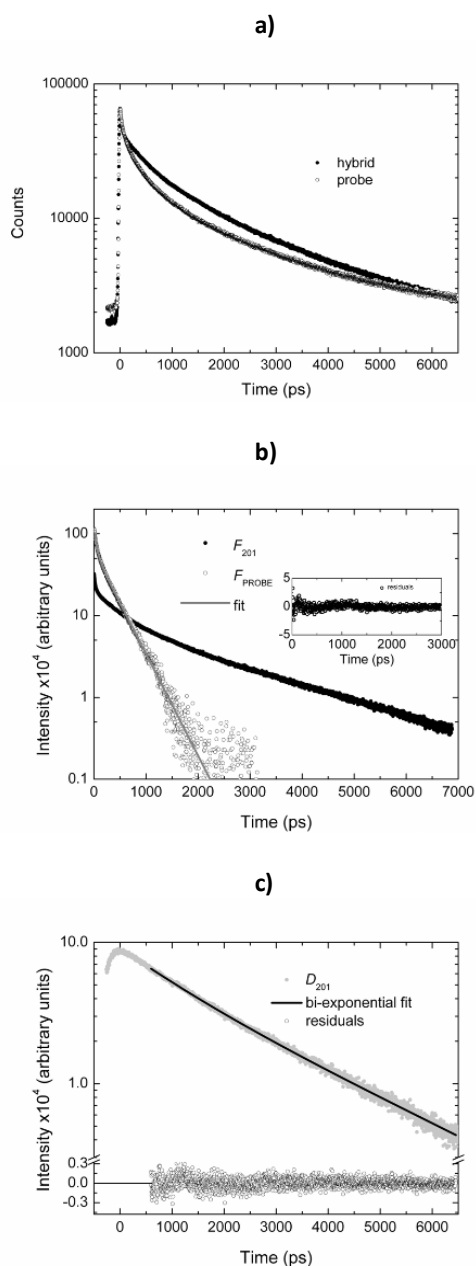

**Figure 4**
Data analysis procedure: a) Rough experimental decay patterns of the samples containing: (black dots) the hybridized DNA bearing DQB1-0201 allelic variant; (open dots) the dual-labelled probe only. b) Integral-normalized fluorescence decays $F_{201}$ (black dots) and $F_{PROBE}$ (open dots) as obtained from the decays in a) upon subtraction of the slow decay component of not-quenched TAMRA. Black line: fitting curve $Y_{PROBE}$ in eq. (1). Inset: residuals of this fit plotted on the same vertical scale as in the main panel. c) Decay $D_{201}$ (grey dots) as obtained by applying eq. (2) to the data in b). Black line: bi-exponential fit with best-fitting parameters $\tau' = 709.92$ ps and $\tau_D = 2724.43$ ps. Open dots: residuals. The arbitrary units in b) and c) are identical to allow the comparison, in amplitude, of $D_{201}$ to $F_{201}$ (see Discussion).

**Table 1** Values of the TAMRA-donor fluorescence decay time, $\tau_D$ ± standard deviation, for the 22-base long dual-labelled DQB1 0201 probe (see **Fig. 2**) bound to: a) synthetic oligonucleotides corresponding to the region between codons 51 and 58 of the listed allelic variants (see **Fig. 1**); b) purified DNAs extracted from HLA-homozygous lymphoblastoid cell lines. NA: sample not available.

| DQB1- allelic variant | a) oligonucleotides $\tau_D$ ± SD (ps) | b) DNAs $\tau_D$ ± SD (ps) |
|---|---|---|
| 0201 | 2725 ± 3 | 2723 ± 3 |
| 0501 | 2514 ± 3 | 2512 ± 5 |
| 0502 | 2404 ± 11 | N.A. |
| 05031 | 2427 ± 2 | N.A. |
| 0602 | 2430 ± 9 | N.A. |
| 0301 | 2481 ± 3 | 2479 ± 3 |
| 0302 | 2599 ± 3 | 2603 ± 5 |
| 0402 | 2462 ± 3 | 2466 ± 5 |